\documentclass[a4paper,fleqn,usenatbib]{mnras}
\usepackage{psfig}

\usepackage{graphicx}	% Including figure files
\usepackage{multirow}
\usepackage{hyperref}
\usepackage{rviewport}

\usepackage{times}
\usepackage{amssymb,amsmath}

\def\lsim{ \lower .75ex\hbox{$\sim$} \llap{\raise .27ex \hbox{$<$}} }
\def\gsim{ \lower .75ex \hbox{$\sim$} \llap{\raise .27ex \hbox{$>$}} }

\title[Multimessenger role of BL Lacs] 
{On the radiation energy density in the jet of high-energy emitting BL Lac objects and its impact on their multi-messenger role}

\author[Tavecchio, Oikonomou \& Righi]
{F. Tavecchio$^1$\thanks{E--mail: fabrizio.tavecchio@inaf.it}, F. Oikonomou$^2$, C. Righi$^1$\\
$^1$INAF -- Osservatorio Astronomico di Brera, via E. Bianchi 46, I--23807
Merate, Italy\\
$^2$European Southern Observatory, Karl-Schwarzschild-Str 2, D-85748 Garching bei Munchen, Germany\\
}

\begin{document}

% \date{Accepted 1988 December 15. Received 1988 December 14; 
% in original form 1988 October 11}

%\pagerange{\pageref{firstpage}--\pageref{lastpage}} \pubyear{2007}

\maketitle

\begin{abstract} 
We examine the potential multi-messenger role of BL Lac objects emitting at high energy (so-called HBL) focusing on the limits on the energy density of soft radiation in the jet frame, a critical parameter which regulates the proton cooling and the fragmentation of ultra-high energy cosmic ray (UHECR) nuclei possibly accelerated in the jet. We show that (under the assumption that the high-energy emission bump is dominated by inverse Compton emission) the energy density of {\it any} external soft radiation field ({\it e.g.} produced by a layer surrounding the jet or in the accretion flow) cannot be larger than few times that associated to the observed synchrotron radiation produced in the emission region. Quite interestingly, the constraint that we derive is generally stronger than the limit obtained from the condition that the source is transparent to very-high energy $\gamma$-rays. Using this constraint we can derive a robust upper limit for the efficiency of the photopion reaction leading to the emission of PeV neutrinos, $f_{\pi}\lesssim 10^{-5}$, which makes HBL quite inefficient neutrino sources. For the photodisintegration of nuclei the results are more dependent on the spectral properties of the radiation field. The photodisintegration efficiency is safely below 1 (and nuclei can escape intact) for a ``canonical" spectrum of the soft radiation field $\propto \nu^{-0.5}$. For radiation fields characterized by a softer spectrum and extended over a large portion of the jet, the efficiency increases and for an appreciable fraction of the sources nuclei with energies above $10^{19}$ eV might suffer significant photodisintegration. 
\end{abstract}


\section{Introduction}

After the detection of an astrophysical neutrino high-energy diffuse emission by IceCube (Aartsen et al. 2013) and the identification of the blazar TXS 0506+056 as a potential neutrino emitter (Aartsen et al. 2018a,b), astroparticle physics fully enters the multimessenger era. However, despite these important leaps, new problems emerged and several old questions still remain unanswered. For instance, it is not clear which class of astrophysical sources (e.g. starburst galaxies, radio-quiet and radio-loud active galactic nuclei) dominates the neutrino emission (see e.g. M{\'e}sz{\'a}ros 2017 for a review). Among the long-standing issues, perhaps the most important concerns the nature of the accelerators of ultra-high-energy cosmic rays (UHECR, see e.g. Kotera \& Olinto 2011, Alves Batista et al. 2019) and the possible connection with the sources of neutrinos.

Characterized by powerful relativistic jets pointed toward the Earth and shining through non-thermal processes, blazars (e.g. Urry \& Padovani 1995; Romero et al. 2017) have been suspected to be the accelerators of UHECRs for a long time (e.g. Rachen \& Biermann 1993, Dermer et al. 2009, Murase et al. 2012, Caprioli 2015) and, most recently,  sources of the high-energy neutrinos (e.g. Murase et al. 2014; Padovani et al. 2014, 2016; Tavecchio \& Ghisellini 2015). In the electromagnetic window blazars show a spectral energy distribution (SED) characterized by two broad non-thermal components, peaking at IR-optical-UV frequencies (depending on the specific source)  and $\gamma$-ray energies, respectively. The first component tracks the synchrotron emission of relativistic electrons, while the high-energy component is often interpreted as inverse Compton (IC) emission from the same energetic electrons (e.g. 
Sikora et al. 1994, Ghisellini et al. 1998).  A hadronic interpretation is also invoked to explain the high-energy emission, either in the form of direct synchrotron emission from protons (e.g. Aharonian 2000) or reprocessed $\gamma$-rays from photo-meson reactions (e.g. Mannheim 1993, B{\"o}ttcher et al. 2013). Mixed (lepto-hadronic) models assume that both leptonic and hadronic emission contribute to the high-energy radiation. 
The detection of the blazar TXS 0506+056 associated with an energetic neutrino has partly clarified the situation. In fact, the shape of the emission of this blazar favors a lepto-hadronic scenario in which the bulk of the $\gamma$-rays are associated to IC emission, while the hadronic component (although vital for the neutrino channel) is thought to provide a rather minor contribution (Ansoldi et al. 2018, Cerruti et al. 2019, Gao et al. 2019, Keivani et al. 2018).

The blazar population presents a division into two broad classes, BL Lac objects and flat spectrum radio quasars (FSRQ). The phenomenological classification is based on the presence (for FSRQ) or the weakness (BL Lac) of broad emission lines in the spectrum, a property possibly related to the nature (efficient {\it vs} inefficient) of the accretion flow onto the central supermassive black hole (e.g. Ghisellini et al. 2009, Righi et al. 2019). FSRQ tend to be more powerful, especially in the $\gamma$-ray band, and to have SED peaks located at low frequencies (Fossati et al. 1998). The positions of the peaks of BL Lacs, on the other hand, span a larger interval of frequencies (Ghisellini et al. 2017). Based on the synchrotron peak frequency BL Lacs are further divided into Low peaked (LBL), intermediate (IBL) and high-peaked (HBL) BL Lacs (Giommi \& Padovani 1994)\footnote{A similar division, generalized to all blazars, among low (LSP), intermediate (ISP) and high (HSP) synchrotron peaked blazars has been introduced by Abdo et al. (2010)}. The latter subclass contains the majority of the sources detected at very-high energy (VHE, above 50 GeV) by Cherenkov arrays. The presence of multi TeV electrons in the jet of HBL -- witnessed by their intense VHE emission -- suggests that these sources host quite efficient acceleration processes. For these reason, HBL are often suggested to be potential accelerators of UHECRs (e.g. Essey et al. 2011, Murase et al. 2012, Rodrigues et al. 2018). Due to the lack of strong thermal features in their optical spectra, it is widely assumed that the high-energy peak of HBL arises from the IC scattering of the synchrotron photons alone (synchrotron self-Compton [SSC] model, e.g. Tavecchio et al. 1998). However, some issues related to the phenomenology of the very-high energy emission of HBL led to the proposal that the radiation field involved in the IC emission could receive  substantial contribution from the synchrotron radiation produced in a slow layer surrounding the faster jet {\it spine} (Ghisellini et al. 2005) or, alternatively, from decelerated portions of the jet (Georganopoulos \& Kazanas 2003). Such schemes can also be adopted to interpret the emission of neutrinos through photomeson reactions (Tavecchio et al. 2014, 2015; Ansoldi et al. 2018). If the emission occurs close enough to the central black hole, the radiation field from the inefficient accretion flow could also play some role (Righi et al. 2019).

Analyses on the potential multimessenger role of blazars are generally based on the simplest one-zone scenario (e.g. Murase et al. 2012, Rodrigues et al. 2018). In this framework, it is assumed that for BL Lacs (and HBL in particular), the only relevant soft radiation field is that associated to the synchrotron component, not considering external sources (i.e. other components of the jet or the accretion flow, as mentioned above). In this paper we intend to re-examine the possible multimessenger role of high-energy emitting BL Lacs (HBL) from a particular perspective, concentrating our study on robust upper limits on the radiation energy density of soft radiation fields in their emission region. Soft photons have a critical role for the potential multimessanger relevance of HBL. On the one hand, too low a density of the targets for the photopion reaction $p\gamma \to \pi X$ represents a strong limitation for the potential production of high-energy neutrinos, since it must be compensated by an exceedingly large cosmic ray luminosity (e.g. Cerruti et al. 2019). On the other hand, the transition to a heavier composition inferred for UHECRs above $\sim 10^{18.5}$~eV (Pierre Auger Collaboration 2014a), likely indicates that  cosmic rays do not suffer effective disintegration by collisions with soft photons in their accelerators and this directly implies that the density of soft targets in the UHECR acceleration sites is relatively low (e.g. Murase et al. 2012, Zhang et al. 2017) and/or the spatial extension of photon fields is limited.  As we will show, the amount of soft radiation in HBL jets (directly constraining the aforementioned processes) can be robustly determined using powerful diagnostics based on the observed SED, {\it regardless of the origin of the (dominant) soft radiation field}. 

In Sect. 2 we describe our diagnostic method and we apply it to derive upper limits on the radiation energy density of high-energy emitting BL Lacs. In Sect. 3 we use the limits to derive the photopion and photodisintegration efficiencies. In Sect. 4 we discuss our results. 

Throughout the paper, the following cosmological  parameters are assumed:
$H_0=70$ km s$^{-1}$ Mpc$^{-1}$, $\Omega_{\rm M}=0.3$, $\Omega_{\Lambda}=0.7$. 
%We  use the notation $Q=Q_X \, 10^X $ in cgs units.

\section{Constraints to radiation energy density in BL Lac jets}

The application of simple but robust emission models allows one to derive the basic physical parameters of the emission region from the observed SED. In the following will first focus on the limits that one can derives from the SED for the energy density of a generic soft radiation field -- intervening in IC scattering and photopion and photodisintegration reactions. The basic assumption behind our treatment will be that, despite the possible importance of hadronic processes, the observed electromagnetic SED is dominated by leptonic emission, i.e. synchrotron and IC radiation of relativistic electrons/pairs . As mentioned above, this is the most used interpretation adopted even by models proposed for TXS 0506+056 (e.g. Ansoldi et al. 2018, Cerruti et al. 2019, Gao et al. 2019, Keivani et al. 2018. Even if this source is not an HBL (being of the LBL/IBL type, or possibly a low-luminosity FSRQ, Padovani et al. 2019) we argue that this result can be generalized to the case under study here. 

Another powerful and quite generic constraint can be extracted considering the opacity of the emission region to the produced high-energy $\gamma$-rays. This constraint is completely independent on the origin and nature of the target field but, besides the density of soft photons, it depends on the spatial extension of the field. Providing a measure of the interaction probability of $\gamma$-rays propagating in the jet, the optical depth is naturally linked to the photodisintegration probability of nuclei.

\subsection{Simple analytical estimates}

\subsubsection{Constraints from the SED}

Even if approximate, an analytical estimate of the relevant physical quantities specifying the jet emission is useful since it helps us to clarify the interconnections between the different parameters and their impact on the inferred value of the radiation energy density.
We closely follow the analytical treatment of Tavecchio \& Ghisellini (2016) (based on Tavecchio et al. 1998) but we extend it to the case of a generic (i.e. not necessarily synchrotron) target photon field for the IC scattering. As in that work we focus the analytical treatment to the case -- usually valid for high-energy emitting BL Lac -- in which the maximum of the high-energy IC peak is determined by scatterings occurring in the Klein-Nishina limit.

The strategy of the calculation is first to derive an estimate of the magnetic field energy density using only the information on the position of the synchrotron and IC peaks in the SED, $\nu_{\rm S}$ and $\nu_{\rm C}$, and then derive the radiation energy density from the ratio between the (total) luminosities of the two peaks, $L_{\rm S}$ and $L_{\rm C}$. 

As commonly assumed, we model the electron energy distribution as a broken power law, with indices $n_1<3$ and $n_2>3$ above and below a break at Lorentz factor $\gamma^{\prime}_{\rm b}$ (primed quantities are measured in the jet frame). In the KN regime, the observed IC peak energy $h\nu_{\rm C}$ does not (strongly) depend on the soft target photon spectrum and is mainly related to the energy of the electrons at the break $\gamma^{\prime}_{\rm b}m_{\rm e}c^2$ by
\begin{equation}
h\nu_{\rm C}=g \gamma_{\rm b}^{\prime} m_{\rm e}c^2 \delta,
\label{nuckn}
\end{equation}
where $g<1$ is a function of the spectral slopes of the synchrotron peak before and after the peak and the spectrum of the target photons and $\delta$ is the relativistic Doppler factor\footnote{$\delta\equiv[\Gamma(1-\beta \cos \theta_{\rm v})]^{-1}$, with $\Gamma$ the bulk Lorentz factor of the jet, $\beta c$ its velocity and $\theta_{\rm v}$ the viewing angle.}. 
From this expression a value for the $\gamma_b$ can be derived:
\begin{equation}
\gamma_{\rm b}^{\prime} = \frac{h\nu_{\rm C}} {m_{\rm e}c^2 \delta g}.
\label{gammab}
\end{equation}
We note that although Eq. \ref{gammab} is strictly valid in the KN regime, if the IC peak is produced in the Thomson regime it can be written as an inequality, since the energy of the electrons emitting at the IC peak would be {\it larger} than that provided by Eq. \ref{nuckn}. 

Inserting Eq. \ref{gammab} into the equation for the observed synchrotron peak frequency, $\nu_{\rm S}= B^{\prime} \gamma_{\rm b}^{\prime 2} \delta e/(2\pi m_{\rm e}c)$ (where $e$ is the electron charge), it is possible  to express the magnetic energy density only as a function of the observed frequencies and the Doppler factor:
\begin{equation}
U^{\prime}_B=\frac{B^{\prime \,2}}{8\pi}=\left( \frac{m_{\rm e}c^2}{h}\right)^4 \frac{g^4 \pi m_{\rm e}^2c^2}{2e^2} \frac{\nu_{\rm S}^2}{\nu_{\rm C}^4} \delta^2.
\label{ub}
\end{equation}
As stressed in Tavecchio \& Ghisellini (2016), $U^{\prime}_B$ depends on the ratio $\nu_{\rm S}^2/\nu_{\rm C}^4$ and, giving the typically large 
distance between the two SED peaks, this generally implies low magnetic energy densities. For instance, using typical values for HBLs, $h\nu_{\rm S}=0.1$ keV, $h\nu_{\rm C}=50$ GeV and $\delta=15$, one finds $ U^{\prime}_B\simeq 4\times 10^{-2}$ erg cm$^{-3}$.

%%%%%%%%%%%%%%%%%%%%%%%%%%%%%%%%%%
\begin{figure}
\vspace{-1.76truecm}
\hspace{-2.8truecm}
  \includegraphics[width=.85\textwidth]{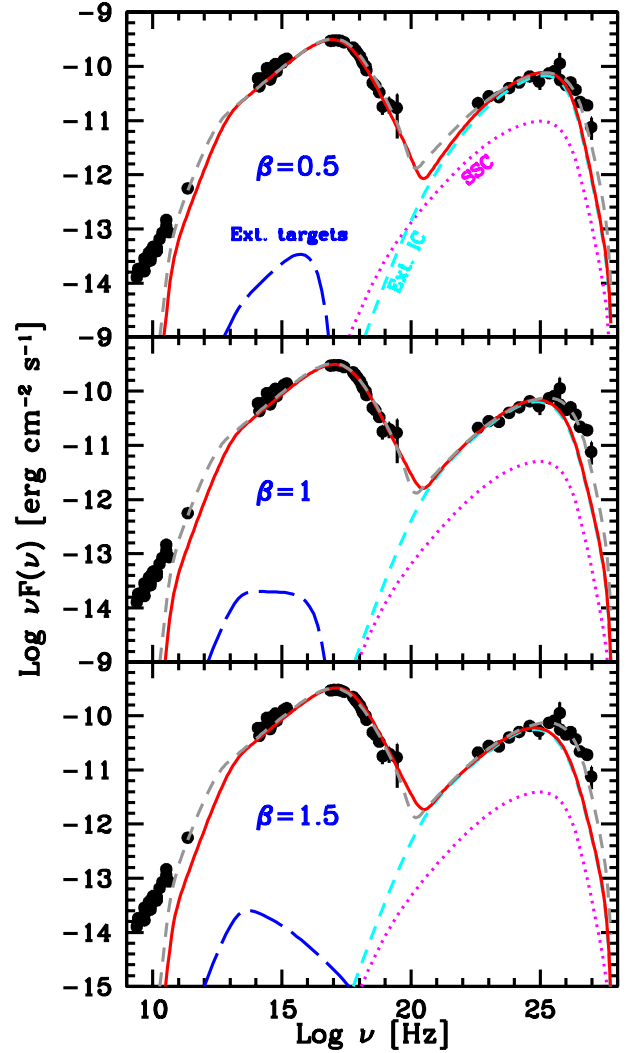} 
  \vspace{-3.4truecm}
  \caption{Spectral energy distribution of Mkn 421 (black filled circles) obtained during the campaign reported in Abdo et al. (2011). 
  The one-zone SSC fit of Tavecchio \& Ghisellini (2016) is reported by the gray dashed line. The solid red lines show different models assuming that the IC by scattering off a dominant external radiation field (cyan dashed line) dominated over SSC (dotted magenta line). In all cases we fix the total radiation energy density of the external field in the jet frame to $U^{\prime}_{\rm ext}=2\times 10^{-3}$ erg cm$^{-3}$ (ten times that derived with the SSC model) and change the slope of the power-law spectrum (long dashed blue line), $\beta=0.5$ (upper panel), 1 (middle) and 1.5 (lower). The blue dashed lines represent the spectra of the external radiation field with arbitrary flux units. In all cases the IC component does not reproduce the high-energy tail of the high-energy peak.}
\label{fig:sed421}
  \end{figure}
%%%%%%%%%%%%%%%%%%%%%%%%%%%%%%%%%%

We now make use of the the ratio between the IC and soft photon luminosity that can be written as:
\begin{equation}
\frac{L_{\rm C}}{L_{\rm S}}=\frac{\xi \, U^{\prime}_{\rm rad}}{U^{\prime}_{B}},
\label{lratio}
\end{equation}
where $U^{\prime}_{\rm rad}$ is the (comoving) total energy density of soft photons and $\xi<1$ is a factor accounting for the reduced efficiency of the IC emission in the KN regime 
(see Tavecchio \& Ghisellini 2016). Therefore we can write the radiation energy density as: 
\begin{equation}
U^{\prime}_{\rm rad} = \frac{1}{\xi}\frac{L_{\rm C}}{L_{\rm S}} \left( \frac{m_{\rm e}c^2}{h}\right)^4 \frac{g^4 \pi m_{\rm e}^2c^2}{2e^2} \frac{\nu_{\rm S}^2}{\nu_{\rm C}^4} \delta^2.
\label{urad}
\end{equation}

We remark that this estimate is extremely generic and provides a constraint that the energy density of {\it any radiation field} (not necessarily that associated to the synchrotron component) must satisfy in order to be compatible with a given SED. We also note that, contrary to naive expectations, it is not possible to allow for larger energy densities by simply using low Doppler factors or smaller radii\footnote{The explicit dependence of the radiation energy density on the observed minimum variability timescale $t_{\rm var}$ could be obtained with the substitution $\delta \to R/ct_{\rm var}$ but this would introduce a new model parameter, the source size $R$.}. As a possible {\it caveat}, we note that a violation of Eq. \ref{urad} could occur if the momentum distribution of the relativistic electrons (especially those with the largest energies) is not isotropic, as recently proposed by Sobacchi \& Lyubarsky (2019). In this case both the actual magnetic and radiation energy densities in the source would be larger than those estimated here through the SED.

\begin{table*}
\centering
\begin{tabular}{lccccccccc}
\hline
\hline
Model    & $\gamma _{\rm min}$ & $\gamma _{\rm b}$& $\gamma _{\rm max}$& $n_1$&$n_2$ &$B$ &$K$ &$R$ 
& $\delta $ \\
\quad [1] & [2]  & [3] & [4] & [5] & [6] & [7] & [8]  & [9] & [10] \\
\hline
1 ($\beta=0.5$)    &$500 $&$ 9.5\times 10^{4} $&$ 2\times10^{6} $&$ 2.2 $&$ 4.9 $&$ 0.22 $&$ 2.3\times 10^{3}$&$ 1 $& 25  \\      
2  ($\beta=1$)  &$500 $&$ 9\times 10^{4} $&$ 2\times 10^{6} $&$ 2.2 $&$ 4.9 $&$ 0.3 $&$ 1.3\times 10^{3}$&$ 1 $& 25 \\   
3  ($\beta=1.5$)  &$500 $&$ 9\times 10^{4} $&$ 2\times 10^{6} $&$ 2.2 $&$ 4.9 $&$ 0.35 $&$ 10^{3}$&$ 1 $& 25 \\    
\hline
\hline
\end{tabular}
\vskip 0.4 true cm
\caption{
Input model parameters for the models of Mkn 421 in Figs. \ref{fig:sed421} and derived magnetic energy density. 
[1]: model.  
[2], [3] and [4]: minimum, break and maximum electron Lorentz factor.  
[5] and [6]: slope of the electron energy distribution below and above $\gamma _b$. 
[7]: magnetic field [G]. 
[8]: normalization of the electron distribution in units of cm$^{-3}$. 
[9]: radius of the emission zone in units of $10^{16}$ cm. 
[10]: Doppler factor. }
\label{tableparam}
\end{table*}

\subsubsection{Optical depth}
\label{subsec:optical_depth}
Gamma rays propagating within a low energy radiation field are effectively absorbed via the conversion into an electron positron pair through the reaction $\gamma\gamma\to e^{\pm}$. The fact that most HBL have been observed at VHE energies by Cherenkov arrays indicates that the sources are generally transparent for such energetic photons. This can be translated into a condition limiting both the energy density and the extension of the possible absorbing radiation fields.

For a sufficiently broad and smooth soft radiation spectrum the optical depth at an energy $E^{\prime}$ can be approximated as:
\begin{equation}
    \tau_{\gamma \gamma} (E^{\prime})=\frac{\sigma_T}{5} d\, n^{\prime}_{\rm rad}(\epsilon^{\prime})\epsilon^{\prime}
\label{eq:tau}
\end{equation}
(e.g. Svensson 1987) where $\sigma_T$ is the Thomson cross section, $d$ the distance travelled by $\gamma$-rays and $n^{\prime}_{\rm ph}(\epsilon^{\prime})$ is the number density of photons at the energy $\epsilon^{\prime}=m_e^2c^4/E^{\prime}$ (all quantities are evaluated in the source frame). If $n^{\prime}_{\rm ph}(\epsilon^{\prime})$ is described by a power law $n^{\prime}_{\rm ph}(\epsilon^{\prime})=k\epsilon^{\prime -(\beta+1)}$ with $\beta>0$, the corresponding optical depth will be an increasing function of the $\gamma$-ray energy 
$\tau(E^{\prime})\propto E^{\prime \beta}$.

The same photons that prevent the escape of very high energy $\gamma$-rays are the ones that contribute to the production of neutrinos in photomeson interactions. The optical depth to photomeson interactions, $\tau_{p \gamma}$, relates to $\tau_{\gamma \gamma}$ via, (e.g. Murase et al. 2016), 
\begin{equation}
\tau_{\gamma \gamma}(E^{\prime}_{\gamma,c}) \approx \,1000 \, \tau_{p \gamma}(E^{\prime}_p),   
\end{equation}
where $E^{\prime}_{\gamma,c} \sim 100~{\rm GeV}~(E^{\prime}_p / 60~{\rm PeV})$, $E^{\prime}_p$ is the comoving proton energy and the factor of $\sim 1000$ reflects the ratio of the Thomson and $p\gamma$ cross sections. Thus, as we show in the following sections the requirement of transparency to the observed $> 100$~GeV $\gamma$-rays imposes a complementary constraint to the integral of $U^{\prime}_{\rm rad}$ and to the path length and thus to the optical depth of photomeson and photonuclear interactions. 

As we have already remarked, while the SED diagnostic allows us to constrain $U^{\prime}_{\rm rad}$, the optical depth (and hence the photodisintegration efficiency) depends on the product $U^{\prime}_{\rm rad} \times d$, where $d$ is the spatial extension of the radiation field (Eq.\ref{eq:tau}). In this sense these two constraints are complementary and together they allow us a strong handle on the properties of the underlying radiation field.

\subsection{An illustrative case: Mkn 421}
\label{subsec:mrk421}
As a concrete example, we discuss in the following the case of the prototypical HBL Mkn 421, exploiting the complete SED obtained during an intermediate-low state by Abdo et al. (2011) and reported in Fig.\ref{fig:sed421}. For the calculations we apply the synchrotron--IC code of Maraschi \& Tavecchio (2003), specified by the source size $R$, the Doppler factor $\delta$, the jet comoving frame magnetic field $B^{\prime}$ and the (jet frame) parameters describing the broken power law electron energy distribution $N^{\prime}(\gamma)$: the minimum, the break and the 
maximum Lorentz factors $\gamma_{\rm min}$, $\gamma_{\rm b}$ and $\gamma_{\rm max}$, the  two slopes $n_1$ and $n_2$ and the normalization, $K$ (for simplicity we omit primes for these quantities).

%%%%%%%%%%%%%%%%%%%%%%%%%%%%%%%%%%
\begin{figure}
\vspace{-2truecm}
\hspace{-0.8truecm}
  \includegraphics[width=.55\textwidth]{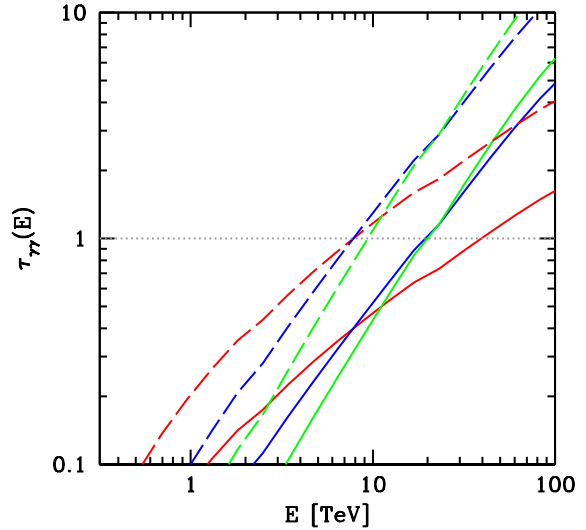} 
  \vspace{-3.8truecm}
  \caption{Optical depth $\tau_{\gamma\gamma}$ as a function of the photon energy (in the observer frame) for the target photon fields assumed for Mkn 421 in Fig. \ref{fig:sed421}. Red: $\beta=0.5$. Blue: $\beta=1$. Green: $\beta=1.5$. Solid lines are calculated assuming that the radiation field has a spatial extension comparable to that of the emission region. For dashed lines, instead, we assume that the radiation field is produced by a layer with extension (layer frame) ten times that of the emission region.}
\label{fig:tau421}
  \end{figure}
%%%%%%%%%%%%%%%%%%%%%%%%%%%%%%%%%%

As stressed in Tavecchio et al. (1998), for the the one-zone SSC model the parameters are uniquely specified once the SED bumps (namely, 
peak frequencies and luminosities) and the variability timescale are well characterized. 
The latter observable is directly linked the value of the source size, through the causality relation 
$R\approx c t_{\rm var}\delta/(1+z)$.  As shown above, the values of the peak frequencies and luminosities can be linked to the other physical parameters, in particular the magnetic and radiation energy densities.

The available data (Fig. \ref{fig:sed421}) provide an excellent description of both the synchrotron and the IC peak. The variability timescale is instead less constrained, since the time scheduling of the campaign was not tailored to probe short variability and it only allows to follow daily--scale variations. For this reason in Tavecchio \& Ghisellini (2016) we performed a SSC modeling for two different variability timescales, $t_{\rm var}=4$ h and 24 h. In Fig. \ref{fig:sed421} we report the case with $t_{\rm var}=4$ h with the dashed gray line. We stress that, despite the ambiguity on the value of the source size, the estimate of the radiation energy density is basically independent of it, since (see Eq. \ref{urad}) it is completely fixed by the SED parameters. In fact, the energy density provided by the SSC modeling with the two variability timescales is in both cases of the order of $U^{\prime}_{\rm rad,SSC}\approx 2\times 10^{-4}$ erg cm$^{-3}$.

To show the impact of a possible source of external photons (coming e.g. from the postulated layer or from the nuclear environment) with a  radiation energy density larger than that associated to the synchrotron component, we show in Fig. \ref{fig:sed421} three models assuming a radiation field characterized by $U^{\prime}_{\rm rad}\simeq 10\times U^{\prime}_{\rm rad,SSC}=2\times 10^{-3}$ erg cm$^{-3}$. To appreciate the role of the spectral slope of the target photon field $\beta$ we calculated the models for hard ($\beta=0.5$), flat ($\beta=1$) and steep ($\beta=1.5$) spectra (blue dashed lines in Fig. \ref{fig:sed421}). The resulting parameters are shown in Table \ref{tableparam}. As can be seen, the impact of the spectral shape is relatively minor. In all three cases the larger radiation energy density forces (because of the observed $L_C/L_s$ ratio, Eq. \ref{lratio})  to increase the magnetic field with respect to the SSC case. In turn, to keep a constant synchrotron peak frequency, one has to decrease $\gamma^{\prime}_{\rm b}$, directly reducing the energy of the IC maximum (Eq.\ref{nuckn}). In all cases, because of the small $\nu_{\rm C}$, the model falls short of the VHE data. In the models reported in Fig.\ref{fig:sed421} we assume that the external radiation field is limited in the frequency range $10^{12}-10^{17}$ Hz (source frame), but we checked that the results are only marginally affected by this choice.

With the same spectral models we have also calculated the optical depth expected at $\gamma$-ray energies. 
We adopt two different geometries of the radiation field. In the first case we assume that the extension of the soft field (measured in the frame of the jet) is limited to the size of the emission region, $d\simeq R$. In the second case, instead, we study a case suitable to model the spine-sheath scenario. We assume that the soft emission is produced by a sheath surrounding the emission region with a length (as measured in the frame moving with the spine) $d=2.5\times R$ (this could correspond to a layer with length ten times that of the spine as measured in its own rest frame and with a relative Lorentz factor $\Gamma_{\rm rel}=4$ between the two zones). In Fig.\ref{fig:tau421} we show the optical depth for $\gamma$-rays as a function of the (observer frame) energy $E$ for the three different slopes of the radiation field. For the calculation we used a convenient approximation of the cross section for isotropic fields (e.g. Coppi \& Blandford 1990). In the most pessimistic case (that assuming the spine-layer configuration, dashed lines) the source starts to became opaque at energies of several TeV, a result compatible with the observed VHE data. Therefore, the constrains on the soft radiation field that we can derive form the opacity is, at least for the specific case of Mkn 421, weaker than that we obtain from the SED fitting.

This example clearly demonstrates that, in agreement with  the analytical treatment, the observed SED allows us to fix a robust {\it upper} limit to the actual radiation energy density in the jet of HBL. In particular, both the shape of the IC peak and the VHE spectrum support the conclusion that, for the specific case of Mkn 421 the assumed $U^{\prime}_{\rm rad}=2\times 10^{-3}$ erg cm$^{-3}$ (i.e. ten times that derived through the SSC model) can be considered a conservative upper limit.

\subsection{Application to a sample of high-energy emitting BL Lacs}

Tavecchio et al. (2010) reported the (non-simultaneous) SED and the physical parameters obtained with the SSC model for 43 BL Lacs whose IC peak is tracked by either {\it Fermi}--LAT or TeV data. In Tavecchio \& Ghisellini (2016) the sample was used to demonstrate that, assuming one-zone emission, these jets are far from equipartition conditions. Here we instead focus on the radiation energy densities associated to those jets.

%%%%%%%%%%%%%%%%%%%%%%%%%%%%%%%%%%
\begin{figure}
\vspace{-1.5truecm}
\hspace{-2.25truecm}
  \includegraphics[width=.66\textwidth]{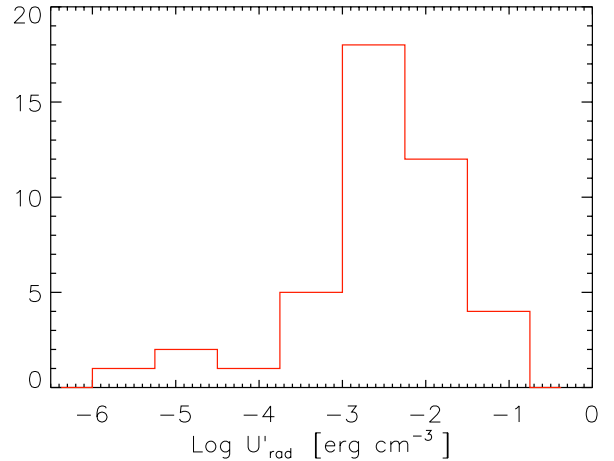} 
  \vspace{-7.truecm}
  \caption{Total (synchrotron) radiation energy density (in the jet frame) evaluated assuming the SSC model for the sample of BL Lac reported in Tavecchio et al. (2010). }
\label{fig:urad}
  \end{figure}
%%%%%%%%%%%%%%%%%%%%%%%%%%%%%%%%%%

In Fig.\ref{fig:urad} we report the distribution of the jet energy densities derived with the SSC model. The majority of the sources have $U^{\prime}_{\rm rad, SSC}<10^{-2}$ erg cm$^{-3}$. The tail at very low values, below $U^{\prime}_{\rm rad, SSC}<10^{-4}$ erg cm$^{-3}$, is associated to the small group of so-called {\it extreme} HBL, characterized by rather peculiar spectral properties (e.g. Costamante et al. 2018).
Generalizing the discussion above for the case of Mkn 421 to the entire sample, we assume that the allowed radiation energy densities, even considering possible contributions by external sources, cannot exceed ten times the value derived through the SSC modeling. Therefore, in the following we assume that $U^{\prime}_{\rm rad}=10\times U^{\prime}_{\rm rad, SSC}$ for all sources.

%%%%%%%%%%%%%%%%%%%%%%%%%%%%%%%%%%
\begin{figure}
\vspace{-1.5truecm}
\hspace{-2.25truecm}
  \includegraphics[width=.66\textwidth]{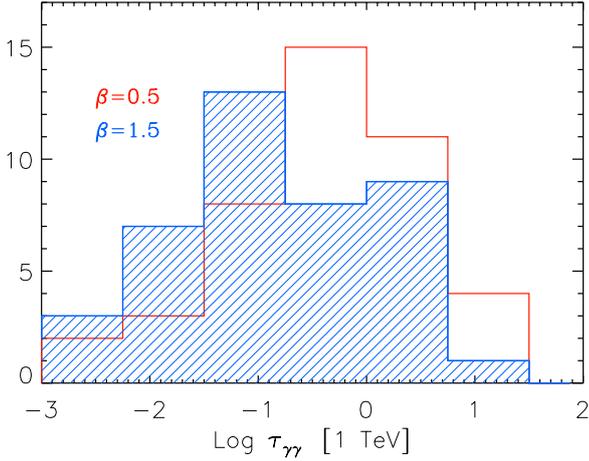} 
  \vspace{-7.truecm}
  \caption{Optical depth at 1 TeV (observer frame) evaluated for the sample of BL Lac reported in Tavecchio et al. (2010) assuming a radiation energy density ten times larger than that derived through the SSC model and a spatially extended external field (see text for details). We show the results for two different values of the spectral slope, $\beta=0.5$ (red) and 1.5 (blue).}
\label{fig:tau}
  \end{figure}
%%%%%%%%%%%%%%%%%%%%%%%%%%%%%%%%%%

In Fig.\ref{fig:tau} we show the distribution of the optical depth for $\gamma$-rays with energy 1 TeV (observer frame) for the sources of the sample. For simplicity we report only the result in the case in which the radiation field extends for a length (as measured in the emitting region frame) 2.5 times that of the emitting region. A fair fraction ($\sim 1/3$) of the sources are characterized by $\tau_{\gamma\gamma}>1$, especially in the case $\beta=0.5$ (red). However, for the majority of the sources the jet is transparent to 1 TeV photons.

\section{Consequences for neutrino and UHECR emission}

As discussed in Sect. 1, the density of the soft radiation in the jet frame is a rather relevant quantity, since it determines the efficiencies of both photopion and photodisintegration reactions. In the following we exploit the limits derived in the previous section to constrain the efficiency of these processes potentially occurring in the jet of HBL.

%%%%%%%%%%%%%%%%%%%%%%%%%%%%%%%%%%
\begin{figure}
\vspace{-2.7truecm}
\hspace{-0.8truecm}
  \includegraphics[width=.6\textwidth]{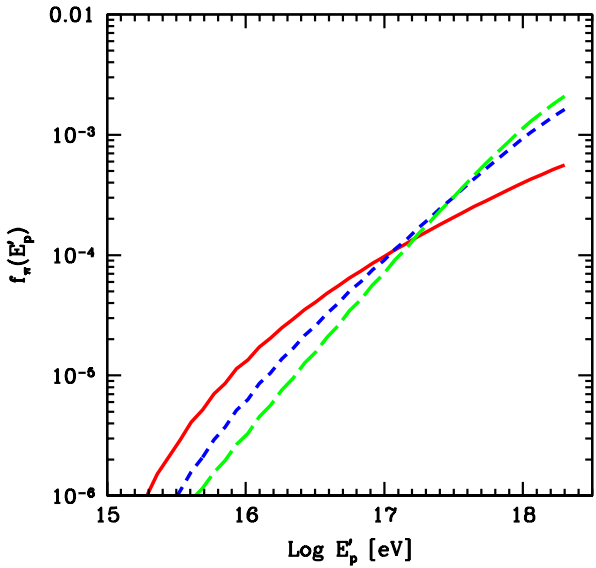} 
  \vspace{-3.4truecm}
  \caption{Efficiency of the photopion reaction, $f_{\pi}$ as a function of the proton energy (in the jet frame) for the target photon fields assumed for Mkn 421 in Fig. \ref{fig:sed421}. Solid red: $\beta=0.5$. Dashed blue: $\beta=1$. Long dashed green: $\beta=1.5$.}
\label{fig:fp421}
  \end{figure}
%%%%%%%%%%%%%%%%%%%%%%%%%%%%%%%%%%

\subsection{Photopion efficiency}

For a proton with energy $E_{\rm p}^ {\prime}$, the efficiency of the photopion reactions is defined as $f_{\pi}(E_{\rm p}^ {\prime})=t_{\rm dyn}^ {\prime}/t_{p\gamma}^ {\prime}(E_{\rm p}^ {\prime})$, in which $t_{\rm dyn}^ {\prime}\approx R/c$ is the dynamical timescale and $t_{p\gamma}^ {\prime}(E_{\rm p}^ {\prime})$ is the photopion cooling time given by:
\begin{equation}
t^{\prime \, -1}_{p\gamma}(E_{\rm p}^{\prime})=c \int_{\epsilon_{\rm th}/2\gamma^{\prime}_{\rm p}} ^{\infty} d\epsilon \frac{n_{\rm t}^ {\prime}(\epsilon)}{2\gamma_{\rm p}^{\prime 2}\epsilon^2} \int_{\epsilon_{\rm th}}^{2\epsilon\gamma_{\rm p}^{\prime}} d{\bar\epsilon}\, \sigma_{p\gamma}({\bar\epsilon})\, K_{p\gamma}({\bar\epsilon}) \,  {\bar\epsilon},
\label{tpg}
\end{equation}
where $\gamma _{\rm p}^{\prime}=E_{\rm p}^{\prime}/m_{\rm p}c^2$, $\sigma_{p\gamma}(\epsilon)$ is the photo-pion cross section, $K_{p\gamma}(\epsilon)$ the inelasticity, $n_{\rm t}^ {\prime}(\epsilon)$ is the number density of targets of energy $\epsilon$ and $\epsilon_{\rm th}$ is given by energy threshold of the process, $\epsilon_{\rm th}=(m_{\pi}^2c^4+2m_{\rm p}m_{\pi}c^4)/4E^{\prime}_{\rm p}$. We evaluate the integrals in Eq. \ref{tpg} using the simple but accurate prescription for $\sigma _{p\gamma}$ and $K_{p\gamma}$ provided in Atoyan \& Dermer (2003), including both single pion (from the $\Delta^+$ resonance) and multi-pion reactions. 

%%%%%%%%%%%%%%%%%%%%%%%%%%%%%%%%%%
\begin{figure}
\vspace{-1.5truecm}
\hspace{-2.25truecm}
  \includegraphics[width=.66\textwidth]{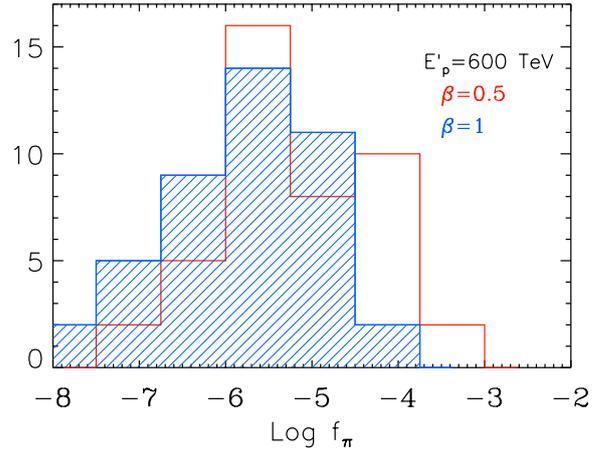} 
  \vspace{-7.truecm}
  \caption{Photomeson efficiency $f_{\pi}$ for protons of energy $E^{\prime}_{\rm p}=600$ TeV for the BL Lacs of the sample of Tavecchio et al. (2010) assuming a radiation energy density ten times larger than that derived through the SSC model and assuming two different spectral slope, $\beta=0.5$ (red) and 1 (blue).}
\label{fig:fpg30}
  \end{figure}
%%%%%%%%%%%%%%%%%%%%%%%%%%%%%%%%%%

%%%%%%%%%%%%%%%%%%%%%%%%%%%%%%%%%%
\begin{figure}
\vspace{-1.5truecm}
\hspace{-2.25truecm}
  \includegraphics[width=.66\textwidth]{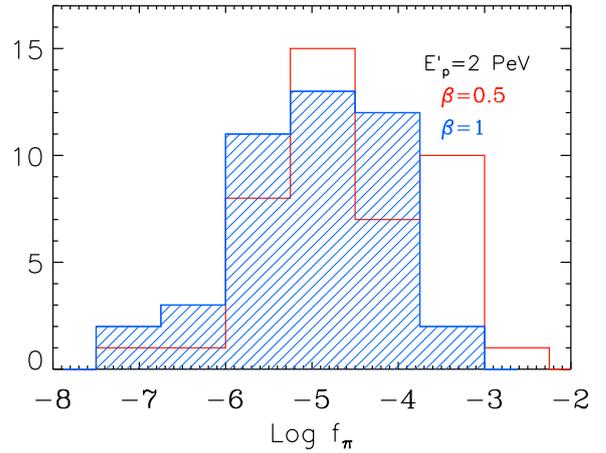} 
  \vspace{-7.truecm}
  \caption{As Fig. \ref{fig:fpg30} but for $E^{\prime}_{\rm p}=2$ PeV}
\label{fig:fpg100}
  \end{figure}
%%%%%%%%%%%%%%%%%%%%%%%%%%%%%%%%%%

In Fig. \ref{fig:fp421} we show the derived value of $f_{\pi}$ as a function of proton energies for the three different spectral models for the soft radiation field used in Fig. \ref{fig:sed421} for Mkn 421. In all cases the efficiency increases monotonically with the proton energy since more and more photons satisfy the threshold condition for the pion production. For the neutrinos detected by IceCube, with energies in the range $E_{\nu}=0.1-5$ PeV, the relevant (jet frame) energies of protons are in the range $E_{\rm p}^{\prime}\simeq E_{\nu} \chi/\delta\approx 0.2-10$ PeV, where $\chi=20$ (e.g. Tavecchio et al. 2014) and we assume $\delta=10$. For this energy range $f_{\pi} \lesssim10^{-5}$. 

In Fig. \ref{fig:fpg30} and \ref{fig:fpg100} we report $f_{\pi}$ for the BL Lac sample assuming, as for Mkn 421, a radiation field with energy density ten times that derived through the SSC model and two different spectral slopes for protons with energy $E^{\prime}_{\rm p}=600$ TeV and 2 PeV. The average value of $f_{\pi}$, as for the case of Mkn 421, is of the order of $10^{-5}$, therefore implying a very small efficiency and, by consequence, a large power for the cosmic ray. 

Specifically, for an observed neutrino luminosity $L_{\nu}$, the required power spent by the jet to energize the cosmic rays (protons for simplicity) is $P_{\rm p}\approx L_{\nu}/f_{\pi}\Gamma^2$ (e.g. Tavecchio \& Ghisellini 2015). For $f_{\pi}=10^{-5}$ and with $\Gamma=10$ we therefore find $P_{\rm p}\approx 10^3\times L_{\nu}$, in agreement with previous estimates based on one-zone models (e.g. Murase et al. 2012). For the HBL to provide a sizeable contribution to the neutrino diffuse background the emitted neutrino luminosities should be $L_{\nu}\simeq 10^{43}-10^{44}$ erg s$^{-1}$ (e.g. Tavecchio et al. 2014, Murase \& Waxman 2016), therefore implying proton power in excess of $P_{\rm p}=10^{46}-10^{47}$ erg s$^{-1}$, about three orders of magnitude larger than that needed to power the observed bolometric electromagnetic output (e.g. Ghisellini et al. 2010, 2014).

\subsection{Photodisintegration opacity}

\begin{figure}
\vspace{-1.5truecm}
\hspace{-2.25truecm}
  \includegraphics[width=.66\textwidth]{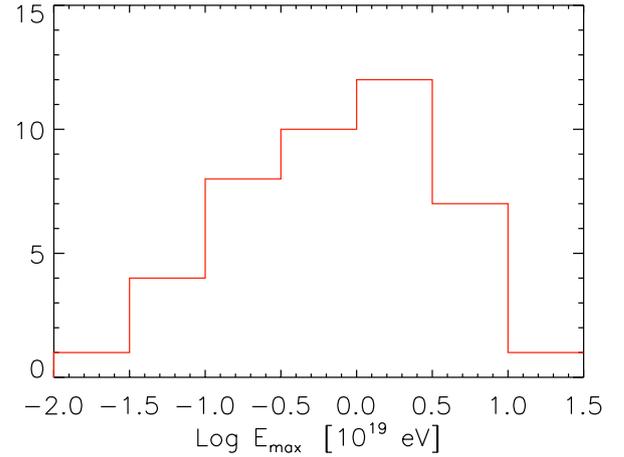} 
  \vspace{-7.truecm}
  \caption{Maximum energy (as measured in the laboratory frame) for protons ($Z=1$) using the Hillas criterium and the physical parameters for the BL Lac jets derived in Tavecchio et al. (2010).}
\label{fig:emax}
  \end{figure}

The maximum energy of protons (in the laboratory frame) for the sources in our sample, $E_{\mathrm{max}}$, obtained by application of the Hillas criterium, is shown in Fig.~\ref{fig:emax}. The average is slightly below $10^{19}$~eV, which means that within the SSC formalism, if these HBLs produce the highest energy cosmic rays, they cannot be protons (see also Murase et al. 2012 and Tavecchio 2014 for the case of extreme HBL). On the other hand, ultra-high-energy (UHE) nuclei could reach such energies. For example, for Mrk\,421 application of the Hillas condition for the parameters listed in Table~\ref{tableparam} gives,
\begin{equation}
E_{Z, \rm max} = 7 \times 10^{20}~\mathrm{eV} \left( \frac{Z}{26} \right) \left(\frac{B}{0.35~\mathrm{G}}\right),
\end{equation}
where we have used the approximation that $\delta \approx \Gamma$. 
\noindent The survival of UHECR nuclei in BL~Lac objects, in interactions with the comoving photon field of the emitting region, has been studied in Murase et al. (2012) and Rodrigues et al. (2018). Here, we focus on the survival of nuclei in the presence of the maximum allowed external target field derived in Sect. 2. 

%%%%%%%%%%%%%%%%%%%%%%%%%%%%%%%%%%
%%%%%%%%%%%%%%%%%%%%%%%%%%%%%%%%%%

%%%%%%%%%%%%%%%%%%%%%%%%%%%%%%%%%%
\begin{figure*}
%\vspace{-1.5truecm}
%\hspace{-2.25truecm}
 \includegraphics[width=.33\textwidth]{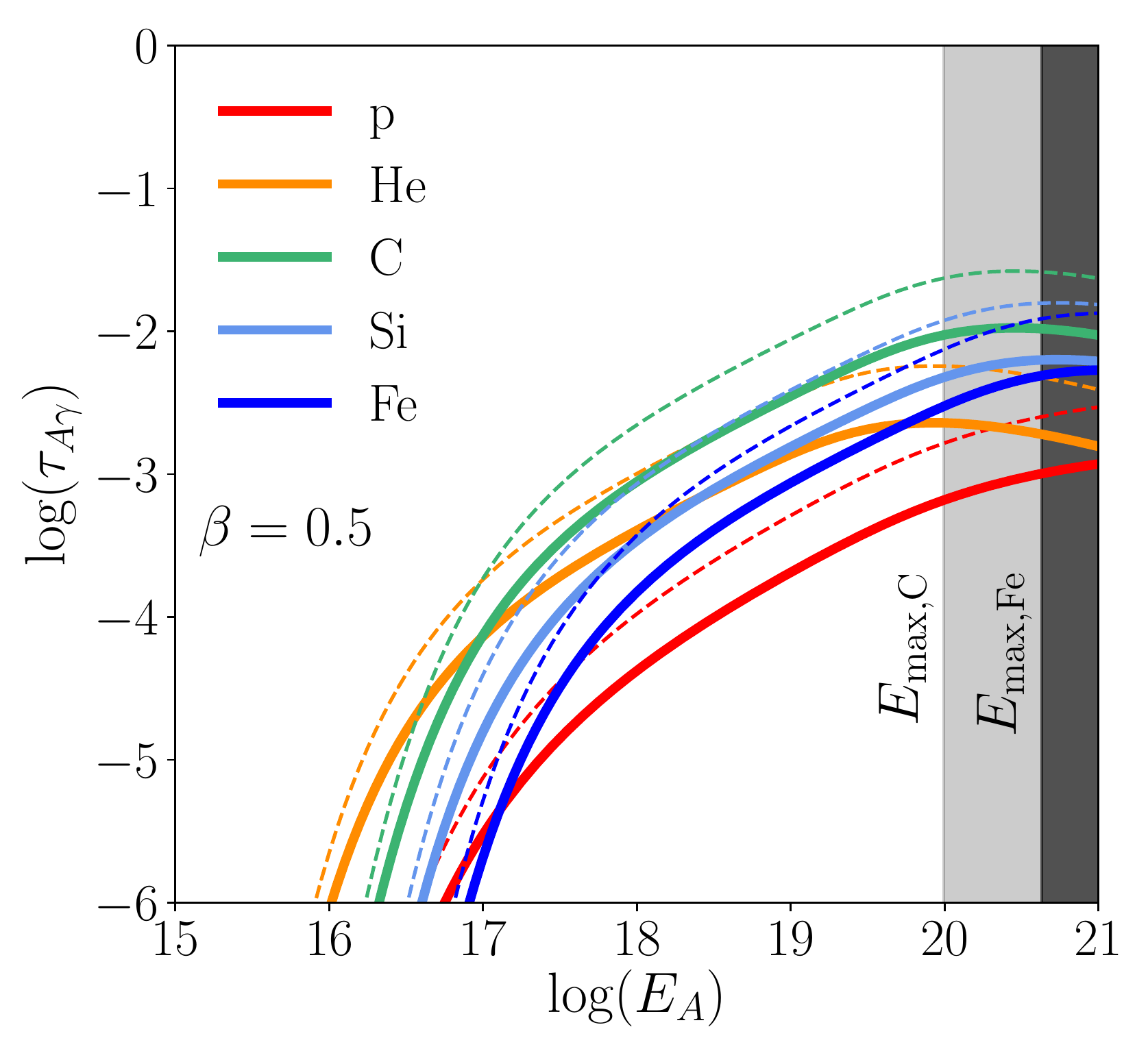}
 \includegraphics[width=.33\textwidth]{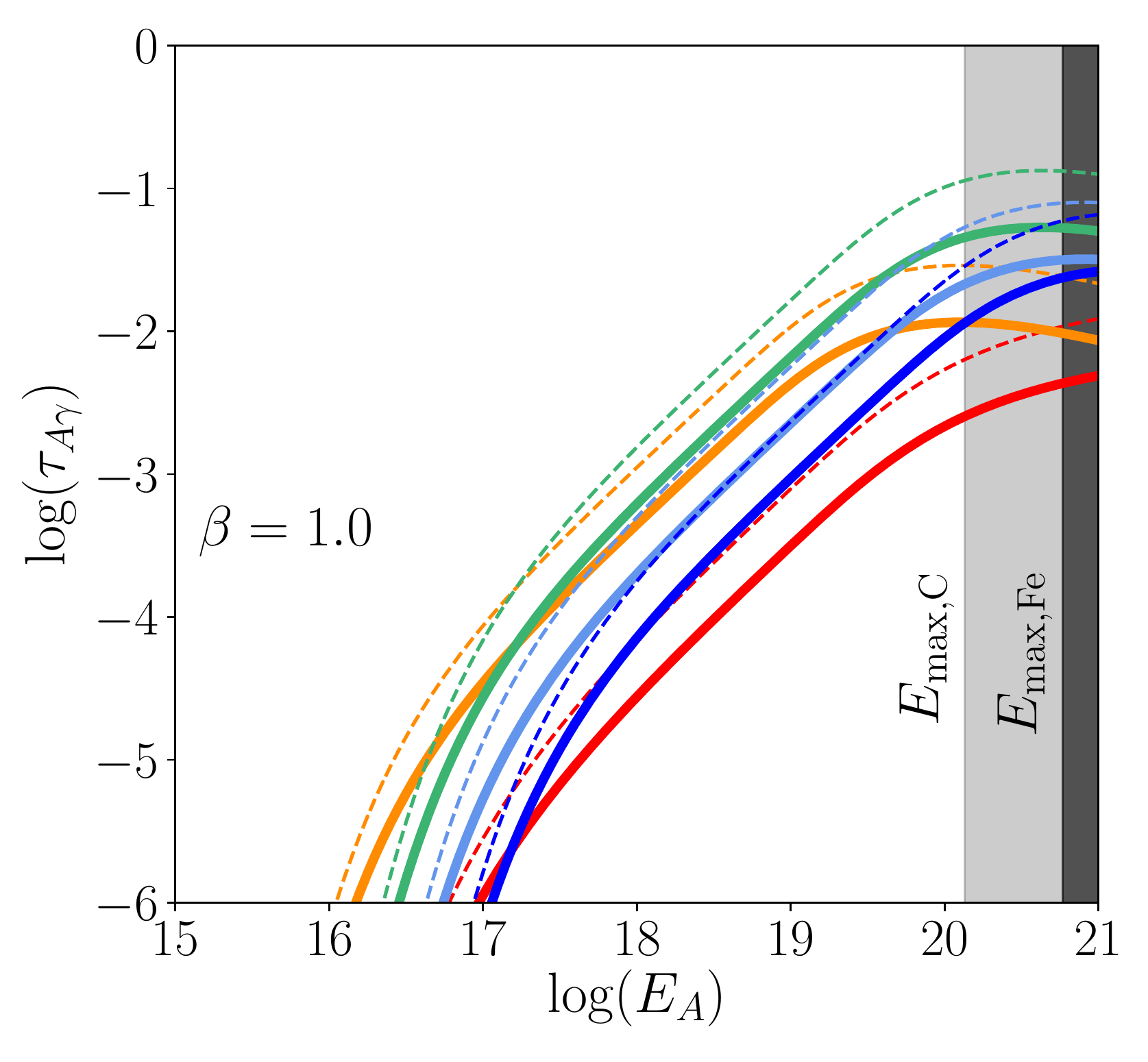}
 \includegraphics[width=.33\textwidth]{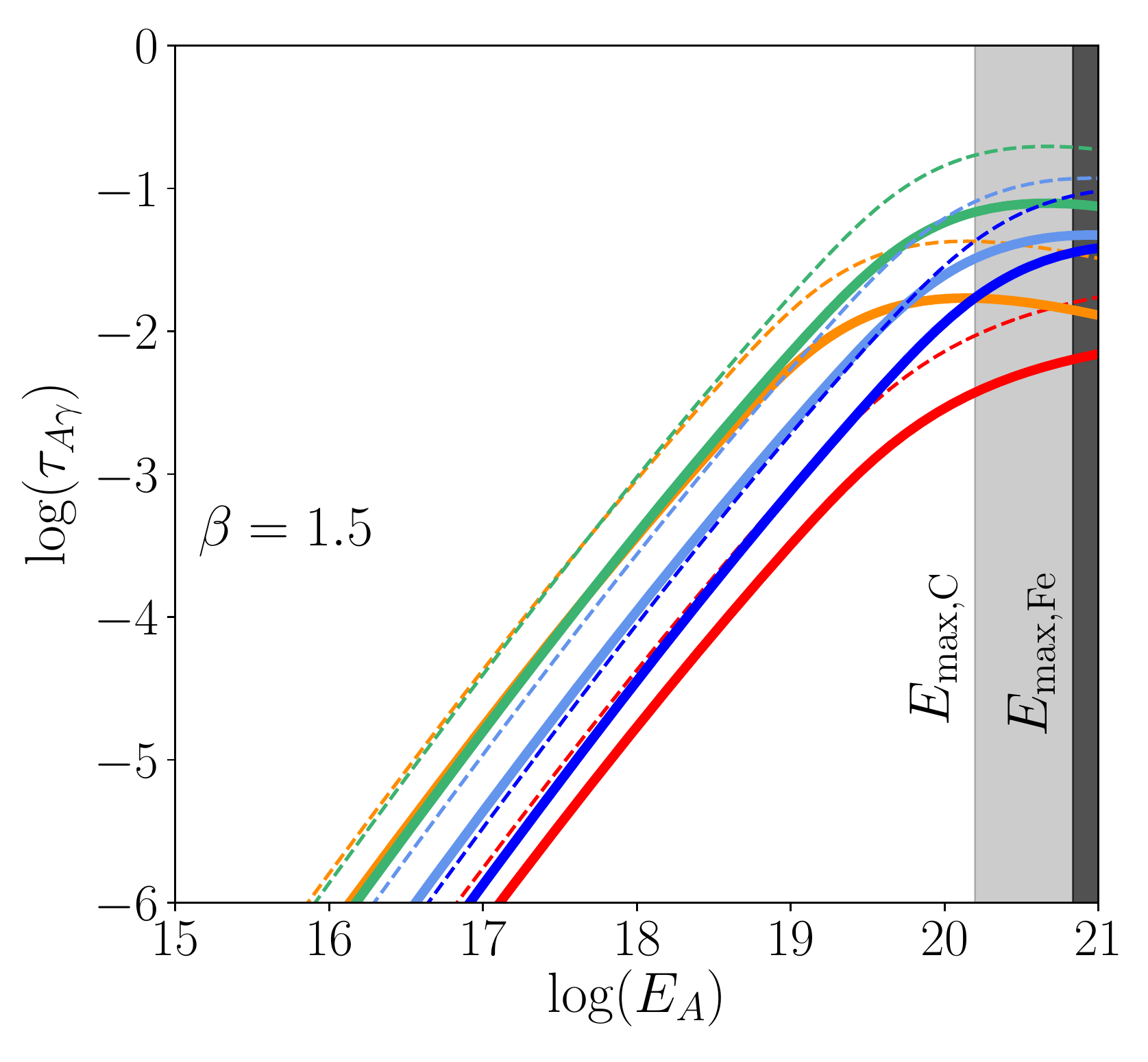}
 % \vspace{-7.truecm}
  \caption{Photodisintegration opacity as a function of the nucleus energy as seen by an observer at redshift $z = 0.03$ in the target photon fields assumed for Mrk\,421, in Fig.~\ref{fig:sed421}, for Helium (He), Carbon (C), Silicon (Si), and Iron (Fe) nuclei. The photopion production efficiency for protons (p) is also shown. Dashed (solid) lines give $\tau_{A\gamma}$ for an assumed external photon field with extension $d = 2.5 \times R$ ($d = R$), where $R$ is the comoving size of the emitting region. The vertical exclusion regions give the energy beyond which Carbon, $E_{\mathrm{max,C}}$ (light grey) and Iron, $E_{\mathrm{max,Fe}}$, (dark grey) cannot be accelerated in Mrk\,421 by application of the Hillas condition using the model parameters listed in Table~\ref{tableparam}.}
\label{fig:disint}
  \end{figure*}
%%%%%%%%%%%%%%%%%%%%%%%%%%%%%%%%%%

The photodisintegration opacity can be expressed in analogy to the $p\gamma$ efficiency, i.e., $\tau_{A\gamma} (E^{\prime}_A)= t'_{\rm esc}/t'_{A\gamma}(E^{\prime}_A)$,
\noindent where $t'_{A\gamma}$ is the photodisintegration cooling time for nuclei with energy $E^{\prime}_A$, and mass number, $A$, and $t'_{\rm esc}$ the timescale beyond which UHECRs stop to interact with photons in the jet. 
 
\noindent The details of the amount of photodisintegration that UHECRs undergo in interactions with the soft target photon field depend on the geometry of the external field, and the way and direction in which UHECRs eventually escape the jet. For definiteness we specialize the calculation to the case in which the external radiation field is supplied by a slower layer surrounding the fast spine jet (Ghisellini et al. 2005). In the absence of precise knowledge of the extension along the jet of the putative sheath (layer) field, and hence $t'_{\rm esc}$ for the escaping UHECRs we consider that the external field might have an extension $R \leq d \leq 2.5 \times R$ in the frame comoving with the spine of the jet as in Sec.~\ref{subsec:mrk421}. We fix the Lorentz factor of the motion of the spine with respect to the sheath field, $\Gamma_{\rm rel} = 4$, for the rest of the photodisintegration discussion. 

We used the photodisintegration cross sections and branching ratios obtained with TALYS 1.6~(Koning et al.), as implemented in~Alves Batista et al.~(2016), to calculate the opacity to photonuclear interactions. For nuclei with mass numbers $A<12$ we used the implementation of~Alves Batista et al.~(2016), based on the parametrizations of~Rachen (1996); Kossov (2002). We have checked that photomeson interactions of nuclei (e.g. Morejon et al. 2019) are not relevant for the maximum energy of nuclei that we expect, in the photon fields we studied, and hence we have not included them. 

In Fig.~\ref{fig:disint}, we show the derived value of $\tau_{A\gamma}$ as a function of the nucleus energy as seen by an observer at redshift $z = 0.03$, for the three different models of the soft radiation field shown in~Fig.~\ref{fig:sed421} for Mrk 421. Four nuclear species are shown, namely Helium, Carbon, Silicon, and Iron. The photopion efficiency is also shown for comparison. Dashed (solid) lines give $\tau_{A\gamma}$ for an assumed external photon field with $d = 2.5 \times R$ ($d = R$).
The vertical exclusion regions give the energy beyond which Carbon, $E_{\mathrm{max,C}}$, (light grey) and Iron, $E_{\mathrm{max,Fe}}$, (dark grey) cannot be accelerated in Mrk\,421 by application of the Hillas condition using the model parameters listed in Table~\ref{tableparam}.The photodisintegration opacity increases with energy but is less than 0.1, independent of the spectral index of the external photon field for all chemical species. We therefore conclude that the jet of Mrk 421 is optically thin to UHECR nuclei, even in the presence of an external radiation field with energy density at the upper limit derived in Sec.~\ref{subsec:mrk421}. 

\noindent
%%%%%%%%%%%%%%%%%%%%%%%%%%%%%%%%%%
\begin{figure*}
  \includegraphics[width=.305\textwidth,clip,rviewport=0 0 0.8 1]{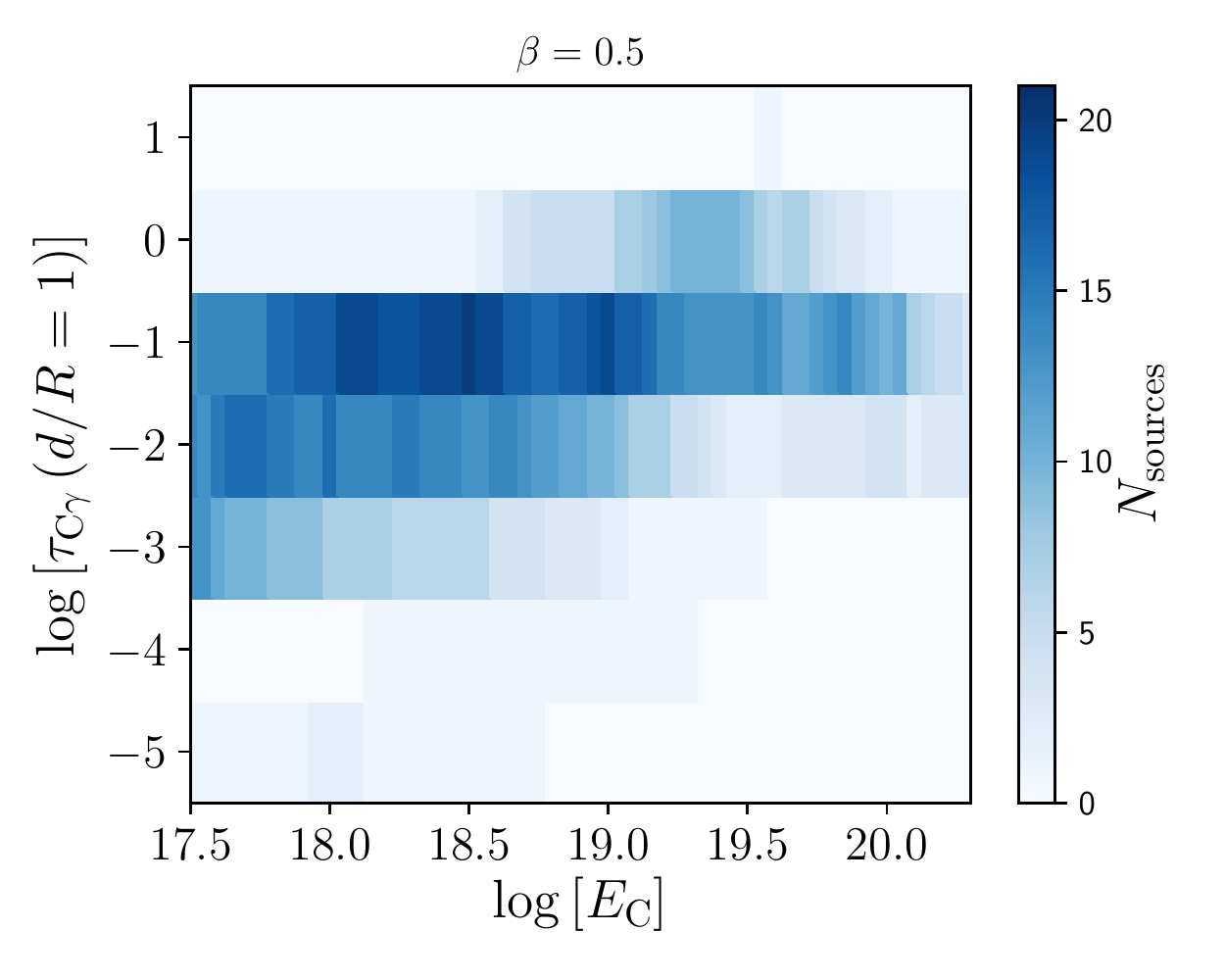} 
    \includegraphics[width=.305\textwidth,clip,rviewport=0 0 0.8 1]{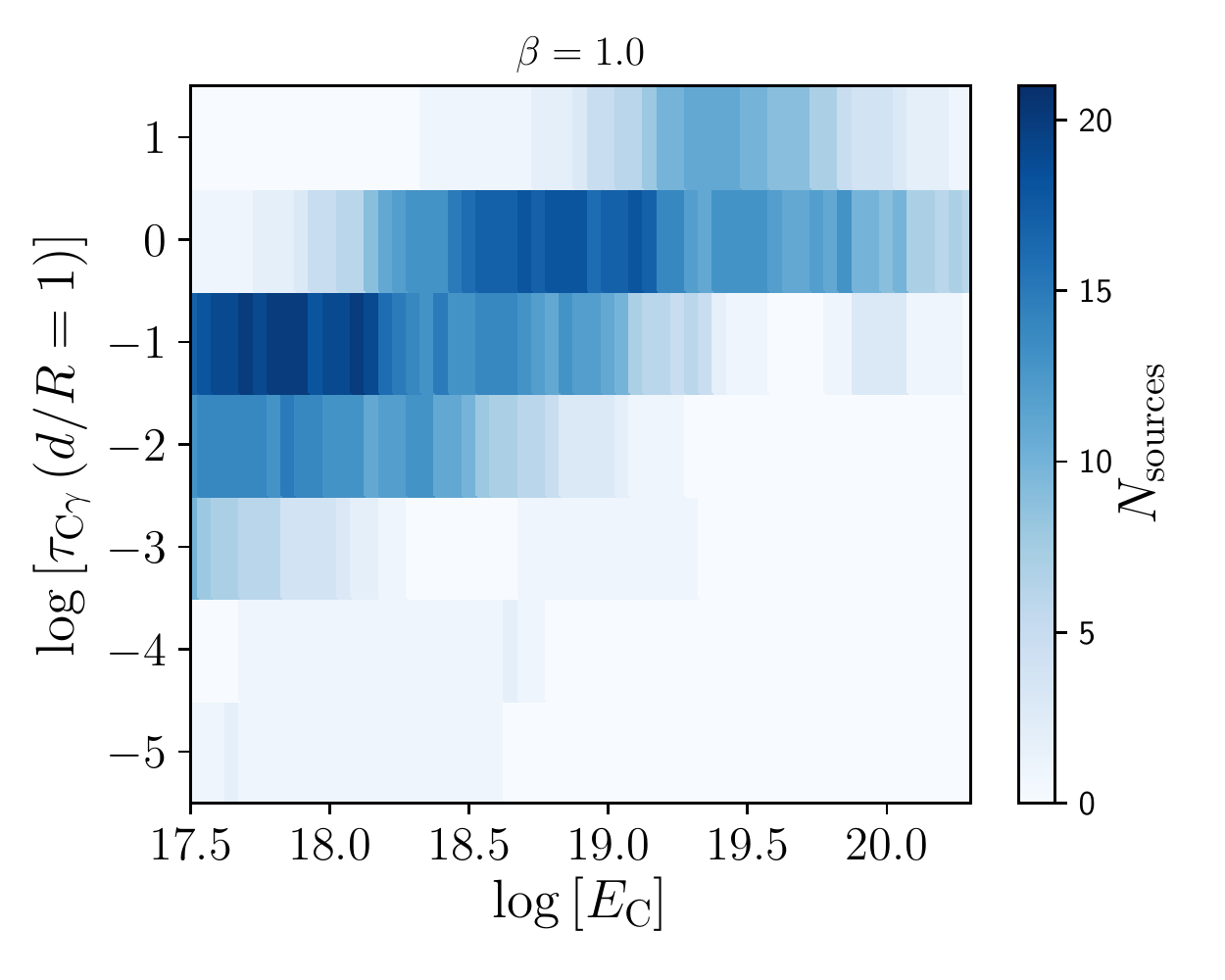} 
      \includegraphics[width=.38\textwidth,rviewport=0 0 1 1]{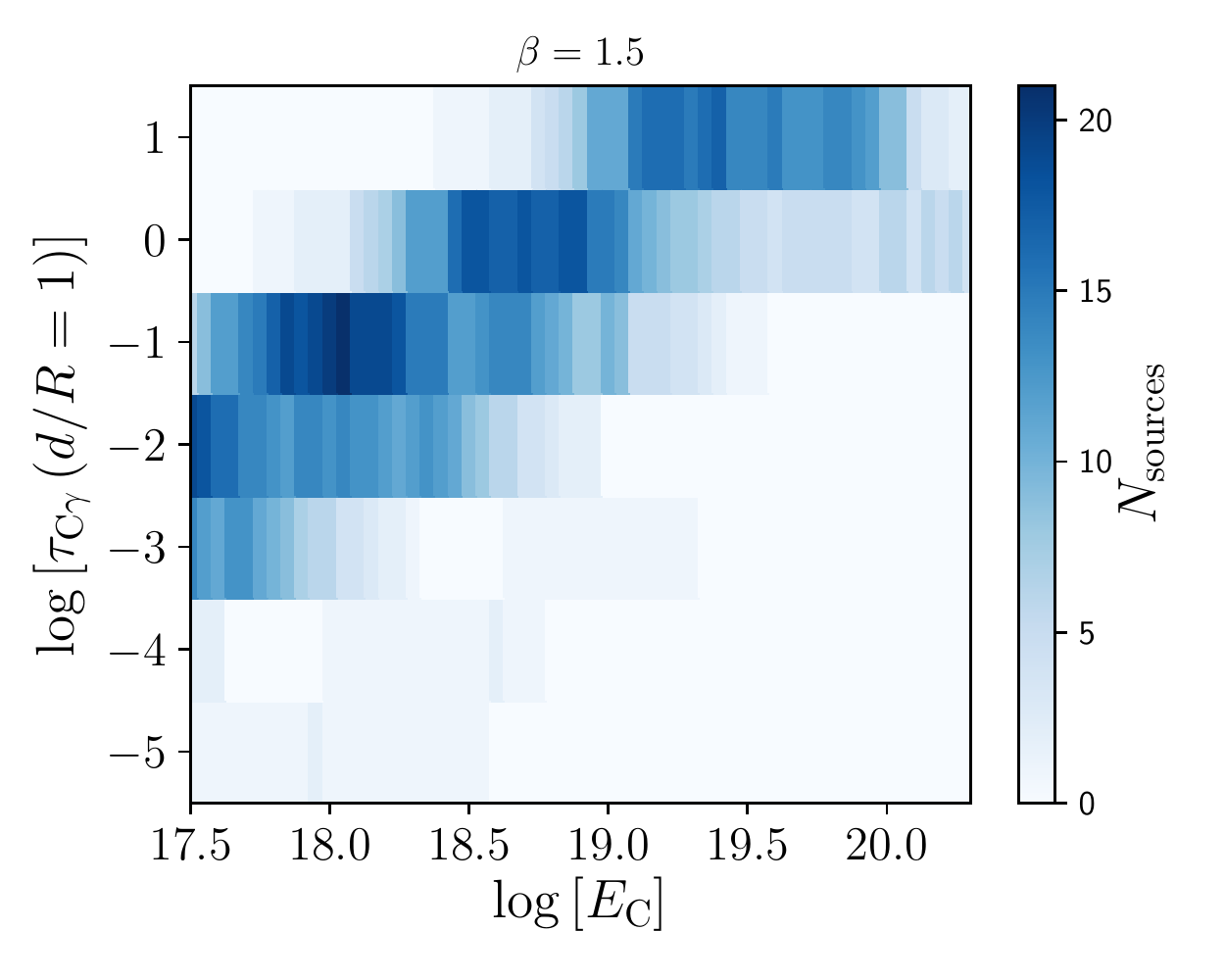}   \includegraphics[width=.305\textwidth,clip,rviewport=0 0 0.8 0.92]{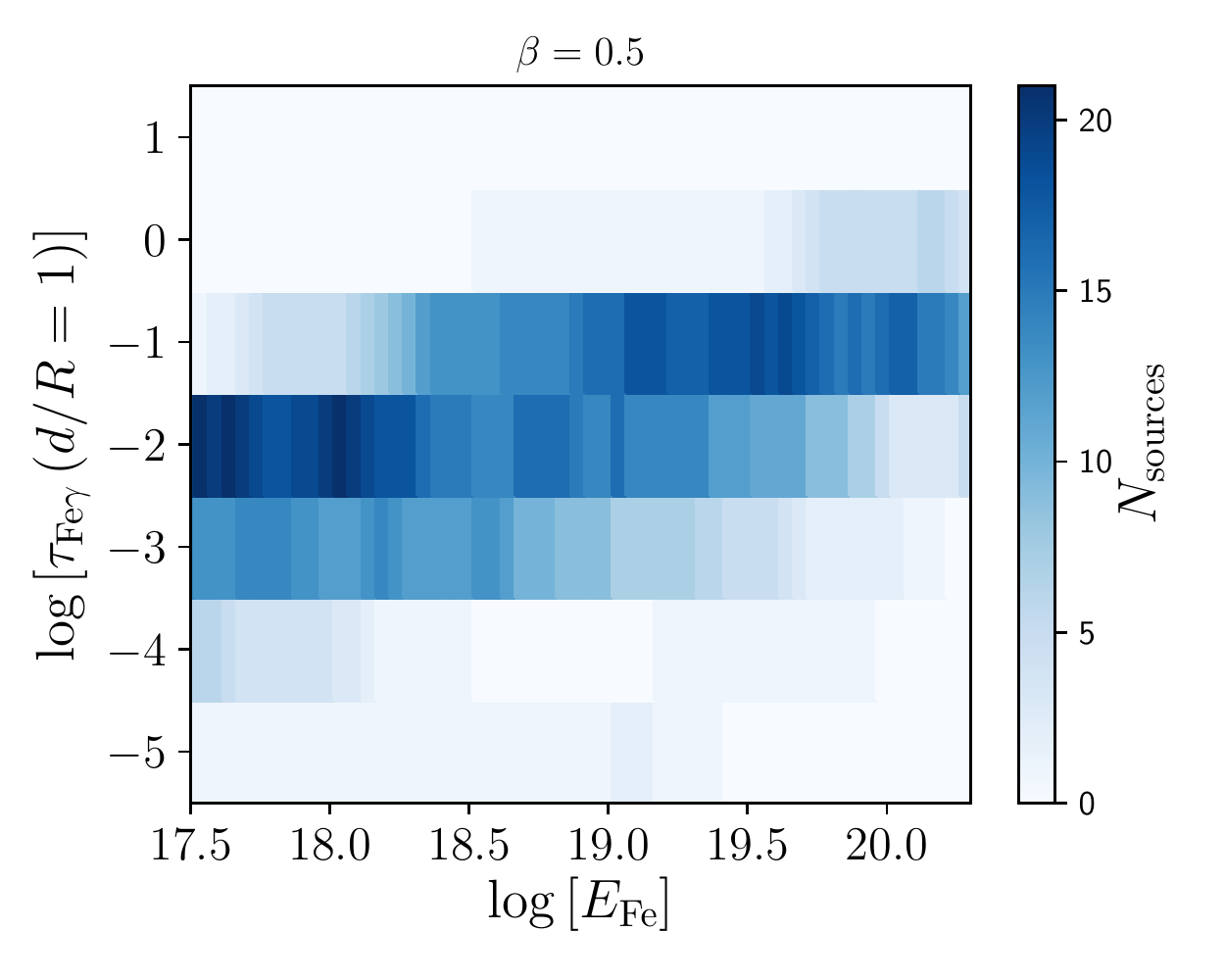} 
    \includegraphics[width=.305\textwidth,clip,rviewport=0 0 0.8 0.92]{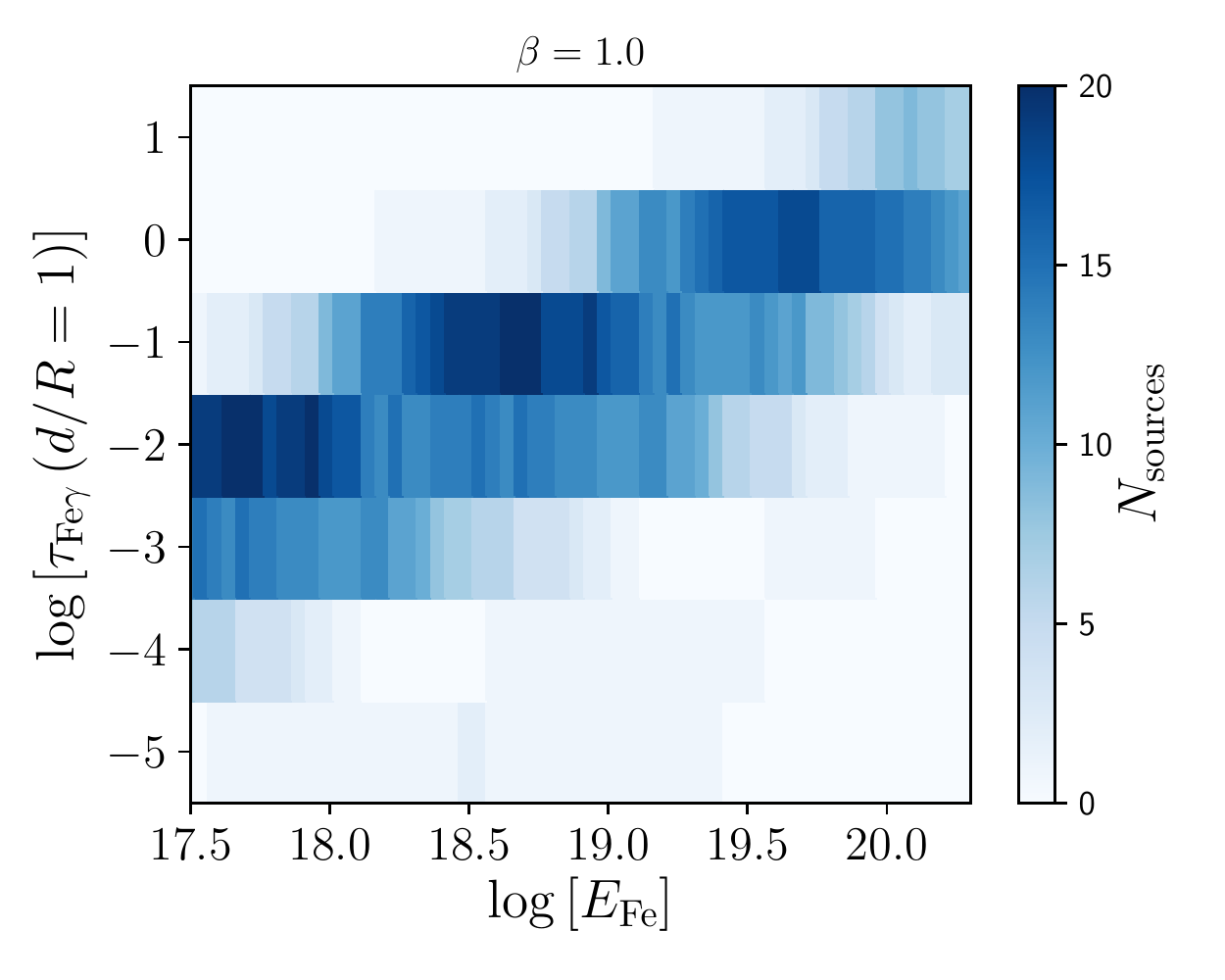} 
      \includegraphics[width=.38\textwidth,clip,rviewport=0 0 1 0.92]{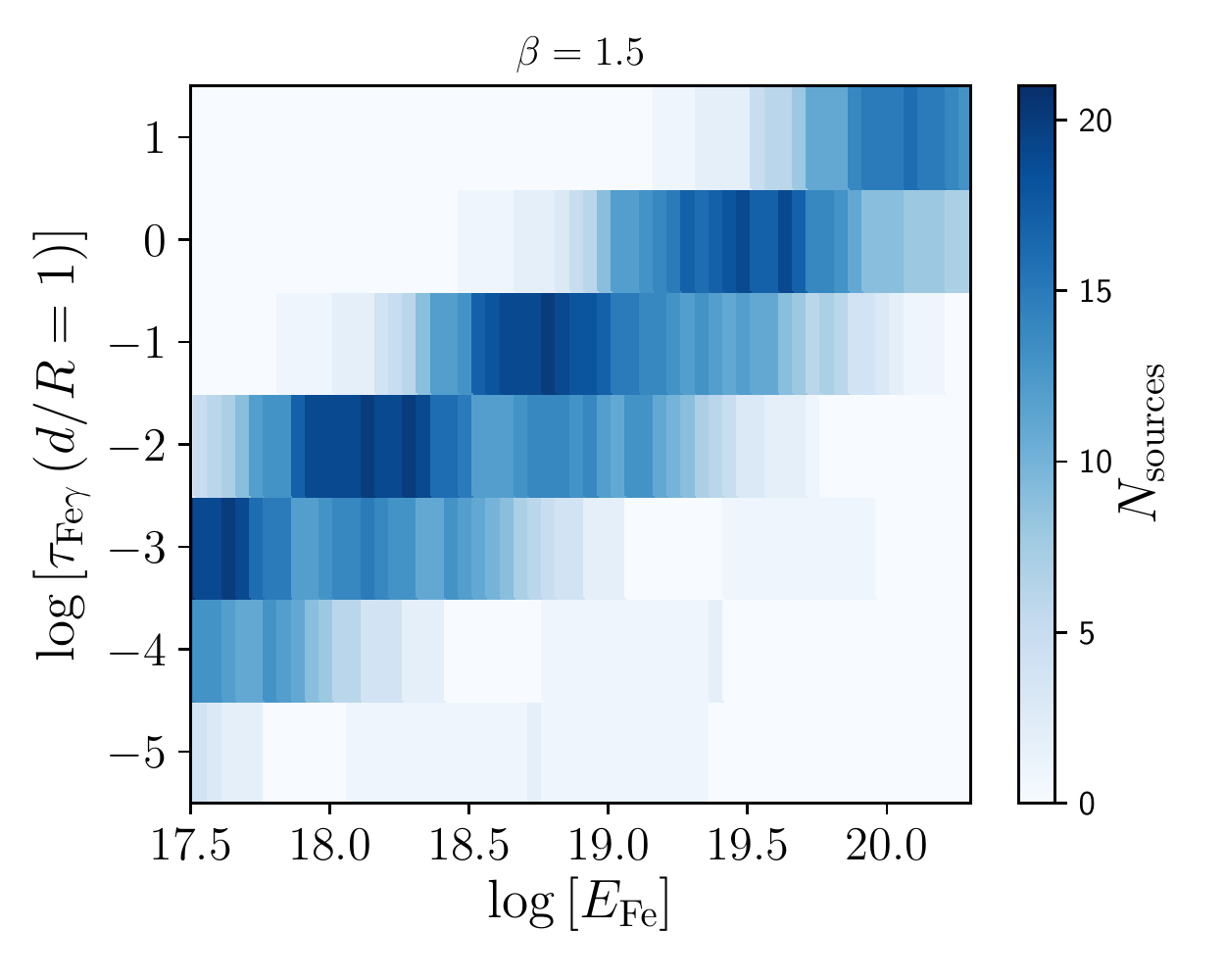} 
  \caption{Photodisintegration opacity as a function of energy in the laboratory frame for Carbon (top row) and Iron (bottom row) nuclei, for the sample of BL Lacs of Tavecchio et al. (2010). From left to right, the assumed external photon field has spectral index $\beta = 0.5, 1.0$, and 1.5. The photodisintegration opacity is given for an assumed external field which extends over $d = R$, where $R$ is the comoving size of the emitting region. The photodisintegration opacity scales up linearly with $d$ for fields that extend over a larger portion of the jet. }
\label{fig:Fe}
  \end{figure*}
%%%%%%%%%%%%%%%%%%%%%%%%%%%%%%%%%%

In Fig.~\ref{fig:Fe} we show the expected photodisintegration opacity for Carbon and Iron nuclei, for the entire BL Lac sample of Tavecchio et al. (2010) as a function of energy. As previously, we assume that $U^{\prime}_{\rm rad} = 10\times U^{\prime}_{\rm rad,SSC}$ as a starting estimate. For each source we show $\tau_{A\gamma}(d/R)$ for a soft photon field with extension equal to the size of the emitting region, $R$, in the frame comoving with the emitting region, up to the maximum energy expected for Carbon and Iron nuclei with the parameters derived in Tavecchio et al. (2010).

 In this case, where $d = R$ the optical depth to photodisintegration at the maximum achievable UHECR energy, $E^{\prime}_{A,\rm max}$, is  $\tau_{A\gamma}(E^{\prime}_{A,\rm max}) < 0.1$ for the majority of sources in the sample when $\beta = 0.5$. 
We observe a trend towards higher opacity when the index of the external photon field is softer. This can be seen in Fig.~\ref{fig:disint} for Mrk\,421 as well. The reason for this trend is that the UHE nuclei typically interact with the low-energy photons of the external photon field which is stronger in the case when $\beta = 1.5$ than for $\beta = 0.5$. This behaviour is the opposite to that observed for TeV $\gamma$-rays which typically interact with the UV part of the external field and thus experience more absorption when interacting with the external field with $\beta = 0.5$. For a field with $\beta = 1.5$ we find that for about half of the sources $\tau_{A\gamma}(E^{\prime}_{A,\rm max}) > 1$. 

The photodisintegration opacity scales up linearly with $d$ for fields that extend over a larger portion of the jet. However, $d$ cannot increase arbitrarily, as the product of $U^{\prime}_{\rm rad} \times d$ is constrained by the optical depth to VHE $\gamma$-rays as we saw in Sec.~\ref{subsec:optical_depth}. In our sample, for twelve (eight) out of 43 sources, a model with $U^{\prime}_{\rm rad} \times d \leq $ 17 (25) $U^{\prime}_{\rm rad,SSC} \times R$ is the maximum allowed for an external photon field with $\beta = 0.5$ (1.5) by this, transparency, condition. We take the requirement for transparency to TeV $\gamma$-rays into account by lowering $U^{\prime}_{\rm rad}$ to keep $\tau_{\gamma \gamma} = 1$ in the calculations that follow. 

Considering a layer field with extension $d = 2.5 \times R$ and $\beta = 0.5$ we find that $\tau_{A\gamma}(E^{\prime}_{A,\rm max}) < 0.1$ for $\sim 50\%$ the sources in the sample (and $\tau_{A\gamma}(E^{\prime}_{A,\rm max}) < 1$ for all but one source), and therefore the majority of jets are expected to be optically thin to ultra-high-energy nuclei even in this case of a very extended external radiation field. Assuming, instead, $\beta = 1.5$, $\tau(E^{\prime}_{A,\rm max}) \gtrsim 1$ for $\sim$half of the sources in the sample (here the fraction of sources with $\tau(E^{\prime}_{A,\rm max}) \gtrsim 1$ does not much change with respect to the case where $d = R$ as the VHE $\gamma$-ray opacity constraint starts to apply for some of the sources). 

We consider a soft photon spectrum with index $\beta = 0.5$ more likely, as it would naturally arise from an underlying electron population with a canonical spectrum with index $-2$. This would imply that HBL jets are in their majority optically thin to UHECR nuclei, but we cannot rule out a external photon field with a softer spectrum, and thus substantial photodisintegration. If photodisintegration does take place, the resulting fragments from $10^{20}$~eV Iron nuclei would have energy per nucleon $E_p \sim E_A/A \sim 10^{18.2}$~eV, and might thus be observable in the energy range where the Pierre Auger Observatory reports the observation of protons~(Pierre Auger Collaboration 2014b), as in the mechanism proposed by~Unger et al. (2015).

\section{Conclusions}

Under the assumption that the high-energy bump of HBL is dominated by IC emission we have derived solid upper limits to the energy density of any radiation field filling the emission region, $U^{\prime}_{\rm rad}$. The constraint on the level of the soft radiation field stems from the strong link between the radiation and magnetic energy density and the positions of the two peaks in the SED. We have derived the upper limit for a sample of BL Lacs already considered in Tavecchio \& Ghisellini (2016) to investigate the relative fraction of particles and magnetic energy in jets. 

The limits on $U^{\prime}_{\rm rad}$ can be used to investigate the efficiency of the photopion and photomeson reactions involving high-energy protons and ultra-high energy nuclei potentially accelerated in HBL jets (e.g. Caprioli et al. 2015). We found rather low efficiency for the photopion reactions, $f_{\pi}\lesssim 10^{-5}$. For all spectral slopes $f_{\pi}$ increases with the proton energy, reaching appreciable (but still low) values $f_{\pi}\simeq 10^{-3}$ only above $E^{\prime}_{\rm p}=10^{18}$ eV. We therefore conclude that it is unlikely that HBL provide a sizeable contribution to the observed neutrino intensity at hundreds of TeV, but their emission could become important at higher energies ($>100$ PeV).

For the photodisintegration of UHECR nuclei the situation is less definite. We studied the transparency of HBL jets to UHECR nuclei under the assumption that cosmic ray acceleration is efficient, i.e. UHECRs can attain the maximum energy allowed by the Hillas criterion. We found that if the external radiation field is characterized by a relatively hard spectrum ($\propto \nu^{-0.5}$) and its emission is concentrated in regions with a size of the same order of that of the region responsible for the observed emission, the photodisintegration optical depth is $\tau_{A\gamma} < 0.1$ and the majority of HBL jets are expected to be optically thin to UHECR nuclei for the chemical species we investigated (Helium, Carbon, Silicon and Iron). On the other hand, if the radiation field is characterised by a soft spectrum, and/or is very extended, photodisintegration can be significant, with $\tau_{A\gamma} \geq 1$ for an appreciable fraction of the sources at the highest attainable energies (approximately half of the sources in the sample we studied). We consider a soft photon spectrum with index $\beta = 0.5$ more likely, as it would naturally arise from an underlying electron population with a canonical spectrum with index $-2$. This would imply that HBL jets are in their majority transparent to UHECR nuclei, but we cannot rule out that in a fraction of HBLs substantial photodisintegration occurs. For the specific case of Mrk 421, the upper limit we derived on $U^{\prime}_{\rm rad}$ is tight and we can rule out 
significant photodisintegration independent of the spectral index of the soft photon field. The survival of UHECR nuclei in HBLs was previously studied by Murase et al. (2012) and Rodrigues et al. (2018) who concluded that HBL jets are transparent to UHECR nuclei in the absence of external radiation fields. Our study is the first to consider the survival of nuclei in the presence of external radiation fields in HBLs and we found that for a sizable fraction of the sources, photodisintegration cannot be ruled out. 

The great majority of the sources we have considered are transparent at VHE even if the radiation energy density is ten times that derived through the SSC model. If we consider an extended external radiation field a fair fraction of the sources have $\tau_{\gamma\gamma}>1$. For most of the sources, however, even in these conditions the optical depth is relatively low and the constraint coming from the SED is stronger than that provided by opacity. The transparency of the jet to VHE photons, implying the possibility that HBL can in principle emit radiation well above 10 TeV, is quite relevant in view of their possible  use as probes of the extragalactic background light or possible anomalies of the transparency (e.g. Biteau \& Williams 2015). Observations at energies above 10 TeV, possible with the upcoming Cherenkov Telescope Array (CTA Consortium 2019), will be quite important to assess the level of opacity and, in turn, the amount of soft radiation in a way complementary to that explored here.

In this work we focused on the limits derived for HBL. Of course, in view of the association of TXS 0505+056 with a high-energy neutrino detected by IceCube it would be important extend the analysis to the other BL Lacs subclasses. We leave a more detailed discussion for a future paper. We only note that, because of the smaller separation between the peaks of the two SED components and the smaller synchrotron peak frequency that characterizes IBL and LBL, substantially larger magnetic fields and, therefore, larger energy densities are compatible with the SED, allowing for larger photopion and photodisintegration efficiencies (for instance in the spine-layer model of Ansoldi et al. 2018 the radiation energy density is $\approx0.3$ erg cm$^{-3}$). Moreover, the limits from the optical depth are less strong than for HBL, since the emission only rarely extends in the TeV band.

Finally we note that our analysis is based on the emission of HBL during quiscent (Mkn 421) or average states (the sample). However, HBL display intense variability, with extended periods of activity marked by high luminosity at all frequencies. While it is conceivable that the physical properties of the emission regions during these active phases could change drastically with respect to those characterizing the low level of activity, we consider it unlikely that the radiation energy density can decrease substantially.

\section*{Acknowledgments}
We thank the referee for constructive comments. FT acknowledges  contribution from the grant INAF CTA--SKA ``Probing particle acceleration and $\gamma$-ray propagation with CTA and its precursors". FO is supported by the Deutsche Forschungsgemeinschaft through grant 
SFB\,1258 ``Neutrinos and Dark Matter in Astro- and Particle Physics''.

\end{document}